\documentclass{pasj00}
\draft

\begin{document}
\SetRunningHead{Author(s) in page-head}{Running Head}
\Received{2002/03/25}
\Accepted{2002/05/27}

\title{Extended Thermal X-ray Emission from
the Spiral-Dominant Group of Galaxies HCG 57}

\author{Yasushi \textsc{Fukazawa}, Naomi \textsc{Kawano}, Akimitsu
\textsc{Ohto}}
\and
\author{Hirofumi \textsc{Mizushima}}
\affil{Department of Physical Sciences, School of Science,
Hiroshima University, }
\affil{1-3-1 Kagamiyama, Higashi-Hiroshima, Hiroshima 739-8526}
\email{fukazawa@hirax6.hepl.hiroshima-u.ac.jp}


%


\maketitle

\begin{abstract}
We observed a group of galaxies, HCG 57, with ASCA.
Regardless that their member galaxies are dominated by spiral galaxies,
we detected extended thermal X-ray emission that is attributed to 
hot gas with a temperature of $1.04\pm0.10$ keV.
This is the second clear detection of thermal X-ray emission from
a spiral-dominant group of galaxies after HCG 92.
The luminosity of the thermal emission 
is about $5\times10^{41}$ erg s$^{-1}$ in the 0.5--10 keV
band, which is higher than that of HCG 92, but 
relatively less luminous among groups of galaxies.
The X-ray emission is extended over several member galaxies, and is thus 
associated with the group rather than an individual galaxy.
The metal abundance cannot be well constrained with a lower limit of 0.08 
solar.
The gas-to-stellar mass ratio is $\sim0.3$.
Although this is relatively low among groups, the hot gas is also a
significant component even in the spiral-dominant group.
We suggest that the X-ray faintness of spiral-dominant groups is due to
 the low surface brightness and somewhat low gas mass, at least in the
 case of HCG 57.
\end{abstract}

\section{Introduction}

Groups of galaxies are the poorest systems of clusters of galaxies.
Since the number of member galaxies is quite small, individual galaxies
could have great effects on the group properties.
Many groups of galaxies have been found to contain a significant amount of
X-ray emitting hot gas with a temperature of $\sim$1 keV 
(Ponman, Bertram 1993; Mulchaey et al. 1996).
ROSAT observations revealed that the 
X-ray luminosity and mass of hot gas scatter
widely with a range of 2 orders of magnitude, 
although their galaxy mass is similar within
a factor of $\sim$5 (Mulchaey et al. 1996, 2000).
Ponman et al. (1996) claimed that this scatter is caused by a steep
relation between the temperature and the X-ray luminosity, 
indicating the action of galactic winds to the hot gas.

The metal abundance is also different with a range of 0.05--1
solar among groups (Fukazawa et al. 1996; Davis et al. 1999),
in contrast to a small scatter of 0.2--0.4 solar for rich clusters of galaxies.
They speculated that the shallow potential cannot bind metal-rich gas
ejected from member galaxies.
This scenario is also supported by the low Si-to-Fe abundance ratio of
groups, as compared with rich clusters (Fukazawa et al. 1998).
However, we have not yet understood the cause of significant scatters in metal
abundances and in the hot gas-to-stellar mass ratio.
In order to resolve this issue, further X-ray investigations of groups of 
galaxies are needed.
It has been reported that some correlations exist among the 
physical parameters of 
groups, such as the galaxy velocity dispersion, optical luminosity, X-ray
luminosity, X-ray temperature, and so on 
(Ponman et al. 1996; Helsdon, Ponman 2000; Helsdon et al. 2001).

In the past observations, the temperature and metal abundance have been 
measured mainly for groups of galaxies whose member galaxies are 
dominated by elliptical ones.
This is due to the X-ray faintness of ones dominated by spiral galaxies
(Mulchaey et al. 1996).
Why are spiral-dominant groups of galaxies X-ray faint? 
How much hot gas is there in such groups?
How high is the metal abundance of hot gas in such groups?
These are important questions to understand concerning how member galaxies 
have a relation with a hot intragroup medium.
In the past, only one spiral-dominant group of galaxies has shown
reasonably bright intergalactic X-ray emission.
This is HCG 92, famous as the Stephan's Quintet, which contains a
significant amount of hot intragroup medium (Sulentic et al. 1995; Awaki
et al. 1997).
The X-ray luminosity is $1\times10^{41}$ erg s$^{-1}$, and the
gas-to-stellar 
mass ratio is $\sim$0.1.
There is another possible evidence of a hot intragroup medium 
in the spiral-dominant
group HCG 16 (Ponman et al. 1996), although its X-ray properties are
not well known.
Since the sample is still too poor for the purpose described above, 
we planned to perform ASCA (Tanaka et al. 1994) 
observations of a spiral-dominant group of galaxies.

HCG 57 is one of the Hickson Compact galaxy groups (Hickson et al. 1988;
Hickson 1993) at a
redshift of 0.0304.
It contains eight original member galaxies in the Hickson Catalog, as
tabulated in tables 1 and 2.
For a comparison, an X-ray bright elliptical-dominant galaxy group, HCG 51 
(Fukazawa et al. 1996), is also shown.
It can be seen that these two groups exhibit quite similar properties, 
such as redshift, galaxy velocity dispersion, galaxy density, optical
luminosity, galaxy number, and group size.
The most different feature is a morphology of dominant member galaxies.
The brightest three galaxies of HCG 57 are spiral, while those of HCG 51
are elliptical or S0.
Therefore, a comparison of the X-ray properties between HCG 57 and HCG 51 is
valuable when considering the relation between the member galaxy morphology and
the hot gas in groups.
Ponman et al. (1996) reported, based on a short serendipitous ROSAT
pointed observation,
that HCG 57 is an X-ray emitter with a luminosity of $10^{41.98}$ erg
s$^{-1}$ and a temperature of $0.8\pm0.2$ keV.
Imaging information is, however, not available, possibly due to the poor
data quality; therefore, the existence of a hot intragroup medium is
not known.
We are especially interested in how much X-ray hot gas exists in HCG 57,
because HCG 51 was found to contain a large amount ot hot gas whose mass
is comparable to the total stellar mass in member galaxies.
Note that the galaxy velocity dispersion of HCG 57 is high enough to bind
hot gas gravitationally.
This paper reports on the ASCA detection of X-ray emitting hot gas from HCG 57.
This is the second detailed X-ray analysis of such a spiral-dominant galaxy
group after HCG 92.
We employed 90\% confidence limits throughout this work, 
and used a Hubble constant of 50 km s$^{-1}$ Mpc$^{-1}$.
The solar abundances are taken from Anders and Grevesse (1989).

\section{Observation and Data Reduction}

We observed HCG 57 on 1994 December 23--25, in the ASCA AO-3 phase, 
with effective times of 82 ks and 61 ks for the GIS (Ohashi et
al. 1996; Makishima et al. 1996) and SIS (Burke et al. 1991), respectively. 
The observation mode was the PH normal mode for the GIS and
the 2CCD Faint mode for the SIS.
We accumulated the detected events under the
condition of a telescope viewing direction of $>5^{\circ}$ from the
dark Earth rim, and a magnetic cutoff rigidity of $>8$ GV. 
Moreover, we imposed the condition of a telescope viewing direction 
of $>25^{\circ}$ from the bright Earth rim for the SIS data, 
in order to avoid the effect of light leakage on the CCD chips.
We added all available data from different sensors and chips 
for the GIS and the SIS separately after an appropriate gain
correction. 
The background data were taken from several blank-sky data, 
such as NEP (North Ecliptic Pole), Draco, and QSF 3 fields; 
we then subtracted them from the on-source spectra. 

Figure 1 shows GIS and SIS images of HCG 57 in the 0.5--2.0 keV and
3.5--7.5 keV band.
Significant X-ray emission was detected around the HCG 57 region 
on both the GIS and SIS images in the 0.5--2.0 keV band.
A point-like source was found toward the west of HCG 57; we
could not find any objects around this source using the NED extragalactic data
base.
The source count rates within 5 arcmin of the group center 
are 0.005 count s$^{-1}$ and 0.006 count s$^{-1}$ for the GIS and SIS,
respectively.
The soft X-ray emission extends up to 5 arcmin, centered on the
spiral galaxy NGC 3753, which is the brightest member galaxy,
and cannot be explained by one point source when considering the ASCA
PSF.
Note that the steep decline of the SIS X-ray brightness towards the west is due
to a sensor gap between the CCD chips.
Compared with the ASCA result for HCG 51 (Fukazawa et al. 1996), 
the X-ray emission of HCG 57 is much fainter by an order of
magnitude, regardless of their similar optical properties.
There is marginal evidence of hard X-ray emission around the HCG 57
region on the GIS image, apart from the northern point-like source
located at $\sim5$ arcmin of the group center.
Whether the X-ray source
is point-like or extended cannot be constrained due to poor statistics.
Low efficiency for hard X-ray and short exposure of the SIS makes
it difficult to detect faint hard X-ray emission.

\section{Results}

Since the photon statistics is too poor to resolve spatially, we
accumulated photons within 5 arcmin of the group center in order to make 
the spectrum.
As shown in figure 2, the thus-obtained spectra exhibit a significant Fe-L
line complex at around 1 keV, indicating the existence of hot X-ray gas.
We first tried to fit the spectra with a single Raymond--Smith model 
(Raymond, Smith 1977) with a Galactic photoelectric absorption of
$2.0\times10^{20}$ cm$^{-2}$ (Stark et al. 1992), but a large 
residual remained in the hard band and the reduced $\chi^2$ was 1.69.
Accordingly, we added a power-law model to represent the hard component; 
the fitting results are shown in figure 2.
This model gives an improved fitting with a reduced $\chi^2$ of 1.08, 
as summarized in table 3.
The temperature of the soft thermal component is $1.04_{-0.16}^{+0.09}$
keV, consistent with the ROSAT result.
The metal abundance cannot be constrained with a lower limit of 0.08
solar.
The power-law photon index $\alpha_{\rm ph}$ 
is constrained to be only $1.0_{-1.0}^{+0.9}$.
We fixed the metal abundance and power-law photon index to be 0.3 solar and
1.0, respectively, and thus 
the X-ray flux becomes $3.1\times10^{-13}$ erg s$^{-1}$ 
cm$^{-2}$ in 0.5--10 keV.
The X-ray luminosities of the soft thermal component and the hard power-law
component are $4.2\times10^{41}$ erg s$^{-1}$ (0.5--2 keV) and 
$6.3\times10^{41}$ erg s$^{-1}$ (2--10 keV).
The combined luminosity of the soft and hard components agrees with the
ROSAT result.
The luminosity of the thermal component is only about 10\% of that of the HCG 
51.
The contribution of the northern source to the hard component is
estimated to be at most 30\%.
The origin of the power-law component is unknown,
noting that the 
hard component with a similar luminosity was also detected from HCG 51
(Fukazawa et al. 1996).

In order to estimate the mass of the thermal hot gas, we fitted the
radial profile with the conventional $\beta$ model (Jones, Forman 1984).
The free parameters were $\beta$, the core radius $R_c$, and the central 
gas density $n_0$.
We prepared radial count rate
profiles in the energy range of 1.0--3.0 keV and 0.8--2.0 keV for the
GIS and SIS, respectively, while considering the signal-to-noise ratios.
With the ASCA instrumental full simulator, we modeled the radial count 
rate profile for specific parameters of the $\beta$ model, and compared
it with the real data to minimize the $\chi^2$ value.
A detailed explanation of the fitting is described in Fukazawa (1997).
We performed simultaneous GIS and SIS fittings and obtained 
best-fit parameters of $\beta = 0.40$, $R_c = 3^{\prime}.75$ (199 kpc), 
and $n_0= 3.2\times10^{-4}$ cm$^{-3}$.
Figure 3 shows the best-fit $\beta$ model and data in the radial count
rate profile of the GIS and SIS.
The gas mass and total mass are $M_{\rm gas} = (4.0\pm1.5)\times10^{11} M_{\odot}$ and 
$M_{\rm total} = (6.1\pm1.0)\times10^{12} M_{\odot}$, respectively, within 250 kpc ($4^{\prime}.7$).
Here, we chose this value for the cut-off radius 
because the X-ray emission is not significant over this radius.
The stellar mass in the HCG 57 is derived from the optical blue 
luminosity of each galaxy (table 2),
assuming a mass-to-light ratio of $8 (M_{\odot}/L_{\odot})$ or $3
(M_{\odot}/L_{\odot})$ for early type galaxies (E/S0) or spiral galaxies, 
respectively; and then,  
$M_{\rm stellar}$ becomes $1.2\times10^{12} M_{\odot}$.
Although total mass is somewhat small, it is typical for groups of
galaxies; 
also, the gas-to-stellar mass ratio of 0.3 seems to be relatively small in
comparison with that of X-ray bright groups of galaxies, such as HCG 51, which
has a ratio of 1.0 (Fukazawa et al. 1996).
However, the difference in the cut-off radius between HCG 51 and HCG 57 must
be taken into account.
Due to the flatter gas density profile of HCG 57 than that of HCG 51,
the gas mass could rise significantly.
Therefore, tracing the emission out to a larger radius is necessary to
discuss the exact amount of hot gas.

\section{Discussion}

We observed a spiral-dominant group of galaxies, HCG 57, with ASCA, and
detected extended thermal X-ray emission with a temperature of 1 keV
around HCG 57, together with the hard X-ray component, whose extent 
cannot be explained by one point source.
Soft thermal emission was not associated with the specific galaxy, 
but extended up to 5 arcmin ($\sim$250 kpc).
The luminosity and mass of the X-ray emitting thermal hot gas are
$4.2\times10^{41}$ erg s$^{-1}$ (0.5--10 keV) and $4.0\times10^{11} 
M_{\odot}$, respectively, within 5 arcmin of the group center.
Compared with the temperature and X-ray luminosity relation of groups of
galaxies taken from Xue and Wu (2000) in figure 4, 
the X-ray luminosity of HCG 57 is 
relatively low for a temperature of
1 keV, but still within the scatter.
The ratios of the gas mass to stellar and total mass are 0.3 and 0.068,
respectively, both of which are somewhat smaller than that of rich
clusters and X-ray bright groups.
This gives us a precious detailed X-ray result for the spiral-dominant 
groups of
galaxies after HCG 92, 
and the third significant detection of X-ray emitting hot gas
from such groups.

According to the X-ray properties of the thermal hot gas such as extent,
temperature, mass, and luminosity, we conclude
that it is not an interstellar medium in a specific galaxy, but 
an intragroup medium bound gravitationally by the HCG 57 itself.
The low mass and low luminosity of hot gas is consistent 
with groups whose spiral fraction is large
(Mulchaey et al. 1996).
Our results indicate that the X-ray low luminosity of spiral-dominant
groups is not due to the lack of an intragroup medium for some systems.
For HCG 57, 
the X-ray faintness is mostly attributed to quite a low central
gas density of at most $\sim3\times10^{-4}$ cm$^{-3}$.
This is also the same as in the case of HCG 16 (Ponman et al. 1996).
We think that the high central gas density of HCG 92 is unusual for
spiral-dominant groups because no other such objects are found.
We claim that the amount of hot gas in HCG 57 is not extremely small 
compared with X-ray bright groups.
Therefore, it can be said that the gravitational potential of
HCG 57 is deep enough to bind a significant amount of hot gas.
This property differs from that for individual galaxies; the amount of
hot gas in individual spiral galaxies is much less than that in
individual elliptical galaxies by two or three orders of magnitude.
Then, why are spiral-dominant groups always X-ray faint even if they
contain massive hot gas?
A low gas density of $<10^{-3}$ cm$^{-3}$, as in the case of HCG 57, or
a low gas temperature of $kT<0.5$ keV is thought to be attributed.
It is speculated that the environment of a low gas density makes spiral
galaxies free from ram-pressure stripping, while a high gas density of
bright groups of galaxies and rich clusters strips off the disk
component of spiral galaxies to convert the morphology into S0.

The temperature of 1 keV for hot gas in the HCG 57 is typical for
groups, and sufficiently high to bind a large amount of hot gas.
Such a deep gravitational potential has been mainly found for
elliptical-dominant groups, and HCG 57 also shows a clear evidence of
such a potential for spiral-dominant groups as HCG 92.
This implies that the gravitational potential or dark matter associated
with groups of galaxies does not depend on the type of member galaxies.
Apart from the group-scale gravitational potential, a galaxy-scale potential
must exist, inferred from the double-$\beta$ structure of the X-ray
surface brightness around the group central galaxy
(Ikebe et al. 1996; Mulchaey, Zabludoff 1999).
For HCG 57, the measurement of the gravitational potential associated 
with central spiral galaxies is quite important, because it has not ever
been performed with other methods.
However, we cannot resolve it because of poor spatial resolution; and
observation by Chandra is hopefully awaited.

A significant amount of metals in the hot gas is also found in HCG
57.
Assuming that the metal abundance is 0.2 solar, the iron mass becomes
$2.1\times10^8 M_{\odot}$.
Accordingly, the iron-mass-to-light ratio (IMLR) is $8.0\times10^{-4}$, which
is small by a factor of 10 compared with that of rich clusters of
galaxies.
There are three early-type (E/S0) galaxies in HCG 57, and their
total optical luminosity is $7.8\times10^{10}L_{\odot}$.
Then, an IMLR for only early type galaxies becomes $2.3\times10^{-3}$.
Considering the IMLR of $\sim10^{-2}$ for rich clusters, 
it is possible that the total iron in the hot gas was supplied mainly by 
these early type member galaxies.
Nevertheless, we suggested that spiral member galaxies also contribute
to the iron enrichment of the hot gas because of their dominance in 
HCG 57.
The metal abundance ratio is an essential key to resolve this issue, 
and hence XMM/Newton
and Astro-E2/XRS/XIS observations with good photon statistics will enable us
to measure it.

The hard X-ray component is also detected from HCG 57.
We cannot determine whether this emission is point-like or extended.
If extended, several possible mechanisms can be considered.
One is inverse Compton scattering via cosmic microwave background (CMB)
photons by high-energy electrons in the intragroup space, as suggested
for several rich clusters (e.g. Fusco-Femiano et al. 1999)
and HCG 62 group (Fukazawa et al. 2001; Nakazawa 2001).
The similar X-ray luminosity of the hard component with that of 
HCG62 and HCG51 is also interesting when considering the origin; the
luminosity of the hard component does not depend on the luminosity of X-ray
hot gas.
For example, if the hard component is due to inverse Compton
scattering, high-energy electrons should exist.
Such electrons might be accelerated commonly in groups of galaxies
through galaxy--galaxy or galaxy--plasma interactions (Nakazawa 2001), 
and luminosity of the
hard component is proportional to the amount of electrons not dependent on
their distribution.
On the other hand, the luminosity of the X-ray hot gas strongly depends on the
gas distribution because it is a square measure of the hot gas density.
Therefore, the luminosity ratio of the soft and hard component can differ
from groups to groups, and spiral-dominant groups could exhibit a clear
hard component without being hindered by the soft component.
Anyway, it is important to measure the spatial distribution of
the hard X-ray component in order to consider its origin.
Astro-E2/XIS will enable us such measurements thanks to the large effective
area and low background.

The authors thank T.J. Ponman for many helpful comments.
The authors are also grateful to the ASCA team for
their help in the spacecraft operation and calibration.

\begin{table}[htbp]
\caption{Optical properties of HCG 57 and HCG 51.}
\begin{center}
\begin{tabular}{cccccc}
\hline
\hline
 & Redshift & $\sigma_v^{\star}$ & $l_{\rm sep}^{\dagger}$ & $L_{\rm B}^{\ddagger}$ & $M/L$ 
\\
& & (km s$^{-1}$) & (kpc) & ($L_{\odot}$) & $(M_{\odot}/L_{\odot})$ \\
\hline
HCG 57 & 0.0304 & 269.2 & 72.4 & 10$^{11.42}$ & 69.2  \\
HCG 51 & 0.0258 & 239.9 & 58.9 & 10$^{11.27}$ & 72.4  \\
\hline
\multicolumn{6}{l}{\small $\star$: Galaxy velocity dispersion.} \\
\multicolumn{6}{l}{\small $\dagger$: Averaged galaxy separation.} \\
\multicolumn{6}{l}{\small $\ddagger$: Optical blue luminosity assumed
 $H_0 = 50$ km s$^{-1}$ Mpc$^{-1}$.} \\
\end{tabular}
\end{center}
\end{table}

\begin{table}[htbp]
\caption{Member galaxies in HCG 57, cataloged in the
 Hickson et al. (1988) and Hickson (1993).}
\begin{center}
\begin{tabular}{ccccccccc}
\hline
\hline
Galaxy: & a & b & c & d & e & f & g & h \\
\hline
Name (NGC) & 3753 & 3746 & 3750 & 3754 & 3748 & 3751 & 3745 & -- \\
$\alpha$ (11h m s) & 37 54 & 37 44 & 37 52 & 37 55 & 37 49 & 37 54 & 37 45
 & 37 51 \\
$\delta$ ($21/22^{\circ}$ $'$ $''$) & 58 51 & 00 34 & 58 26 & 59 08 & 01 33
 & 56 10 & 01 15 & 00 43 \\
$m_B$ & 13.99 & 14.32 & 14.63 & 14.51 & 15.37 & 15.22 & 15.84 & 16.75 \\
Type & Sb & SBb & E3 & SBc & S0a & E4 & SB0 & SBb \\
\hline
\end{tabular}
\end{center}
\end{table}

\begin{table}[htbp]
\caption{Results of spectral fittings of HCG57.}
\begin{center}
\begin{tabular}{ccccccc}
\hline
\hline
Model & $kT$ & Abundance & Norm$^a$ & $\alpha_{\rm ph}$ & Norm$^b$ & $\chi^2$/d.o.f. \\
& (keV) & (solar) & & & & \\
\hline
WABS*RS & $1.29_{-0.19}^{+0.84}$ & $0.18_{-0.09}^{+0.22}$ &
 $2.3_{-0.6}^{+0.7}\times10^{-4}$ & & & 1.67 \\
WABS*(RS+PO) & $1.04_{-0.16}^{+0.09}$ & $>0.08$ &
 $1.2_{-1.2}^{+1.3}\times10^{-4}$ & $1.0_{-1.0}^{+0.9}$ &
 $9.5_{-8.1}^{+0.0}\times10^{-6}$ & 1.08 \\
             & $1.03_{-0.11}^{+0.09}$ & $0.3$ (fix) &
 $1.3_{-0.3}^{+0.2}\times10^{-4}$ & $1.0$ (fix) &
 $9.3_{-2.6}^{+2.9}\times10^{-6}$ & 1.03 \\
\hline
\multicolumn{7}{p{13cm}}{Note.}\\
\multicolumn{7}{p{13cm}}{The model WABS, RS, and PO represent the Galactic
 absorption, Raymond--Smith, and power-law model, respectively.} \\
\multicolumn{7}{l}{The column density of the Galactic absorption (WABS)
 is fixed to be $2.0\times10^{20}$ cm$^{-2}$.} \\
\multicolumn{7}{p{13cm}}{The norm$^a$ represents
 a normalization of the Raymond--Smith model in unit of $1.0\times10^{-14}/(4\pi D^2[\mbox{cm}])VEM[\mbox{cm}^{-3}]$ where $D$ is
 a distance to the source and $VEM$ is a volume emission measure of the
 plasma.} \\
\multicolumn{7}{p{13cm}}{The $\alpha_{\rm ph}$ is a photon index of the
 power-law model.}\\
\multicolumn{7}{p{13cm}}{The norm$^b$ is a normalization of the
 power-law model in unit of
count s$^{-1}$ cm$^{-2}$ keV$^{-1}$ at 1 keV.}
\end{tabular}
\end{center}
\end{table}

\begin{figure}[hb]
\FigureFile(40mm,40mm){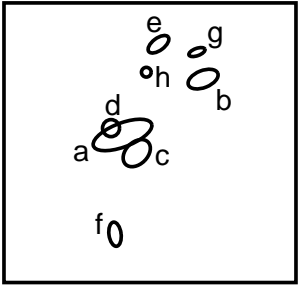}\\
\begin{minipage}[tbhn]{8cm}
\FigureFile(75mm,75mm){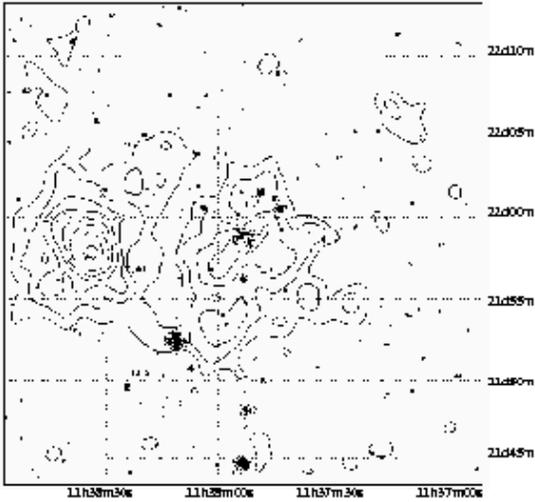}
\end{minipage}\quad
\begin{minipage}[tbhn]{8cm}
\FigureFile(75mm,75mm){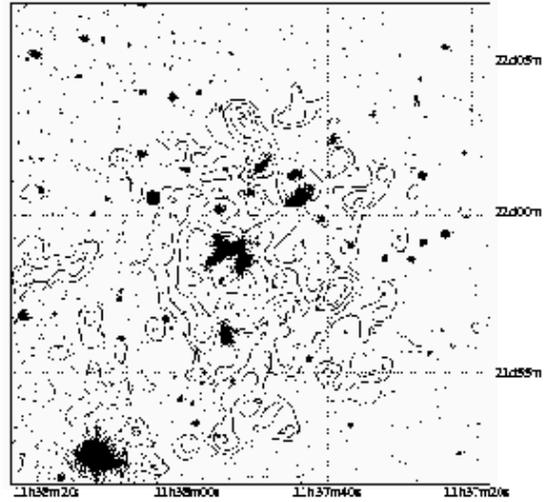}
\end{minipage}
\begin{minipage}[tbhn]{8cm}
\FigureFile(75mm,75mm){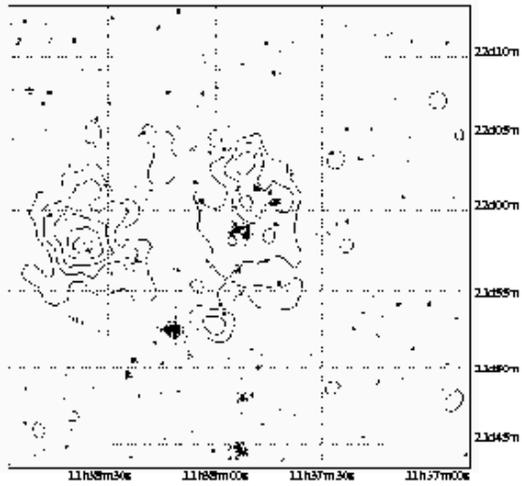}
\end{minipage}\quad
\begin{minipage}[tbhn]{8cm}
\FigureFile(75mm,75mm){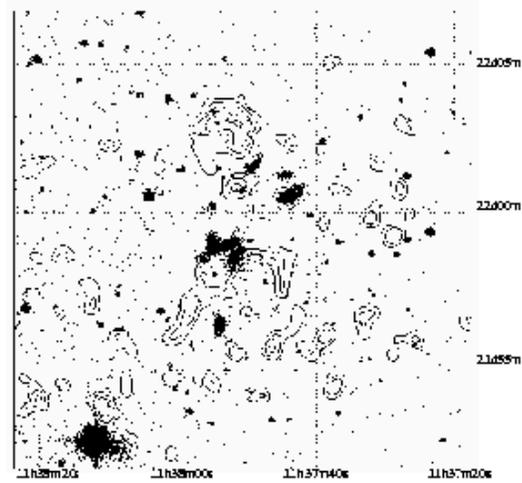}
\end{minipage}
\caption[]{X-ray images (contours) of the HCG 57 obtained with the
 GIS (left) and SIS (right), overlaid on the optical image (from the SkyVIEW
 website, http://skyview.gsfc.nasa.gov/skyview.html). 
The top panel indicates the galaxy position
 (referred to Hickson 1993).
 The middle and bottom panels are images in 0.6--2.0 keV
 and 3.5--7.5 keV, respectively. The contour levels are logarithmically
 spaced by a factor of 1.26.}
\end{figure}

\begin{figure}[hb]
\centerline{\FigureFile(90mm,90mm){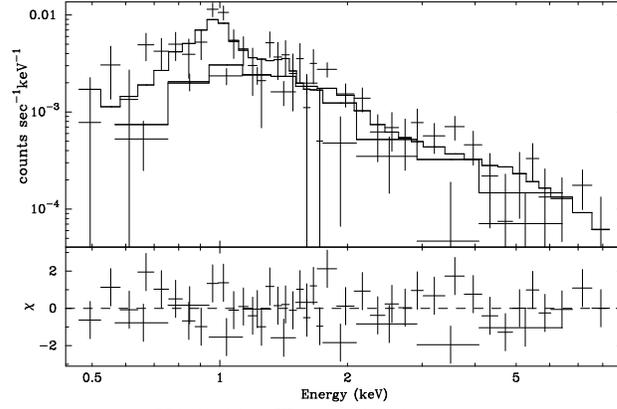}}
\caption[]{GIS and SIS spectra of the HCG 57. The solid lines
 represent the best-fit Raymond-Smith plus powerlaw model.}
\end{figure}

\begin{figure}[hb]
\begin{minipage}[tbhn]{8cm}
\centerline{\FigureFile(95mm,95mm){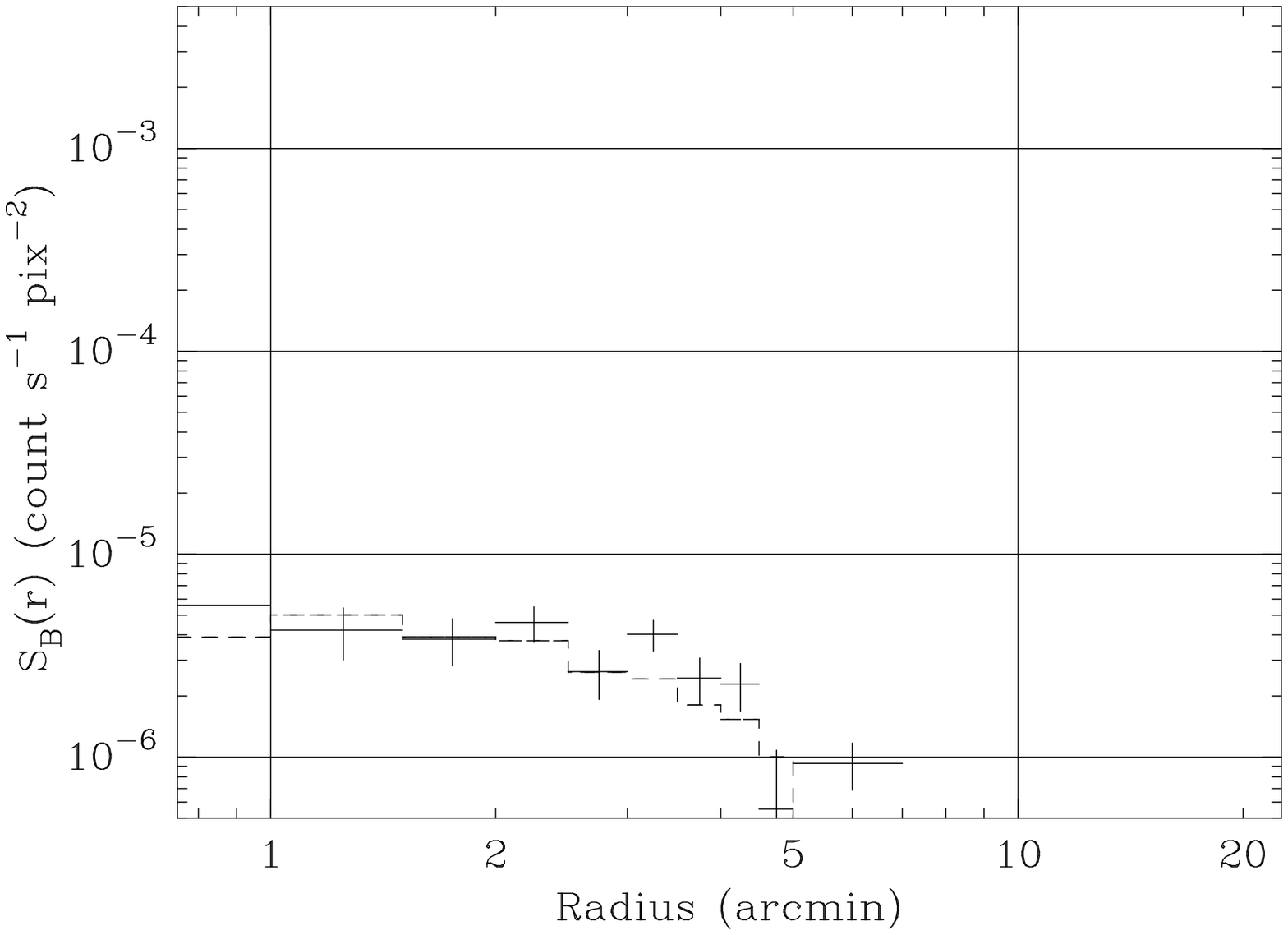}}
\end{minipage}\quad
\begin{minipage}[tbhn]{8cm}
\centerline{\FigureFile(95mm,95mm){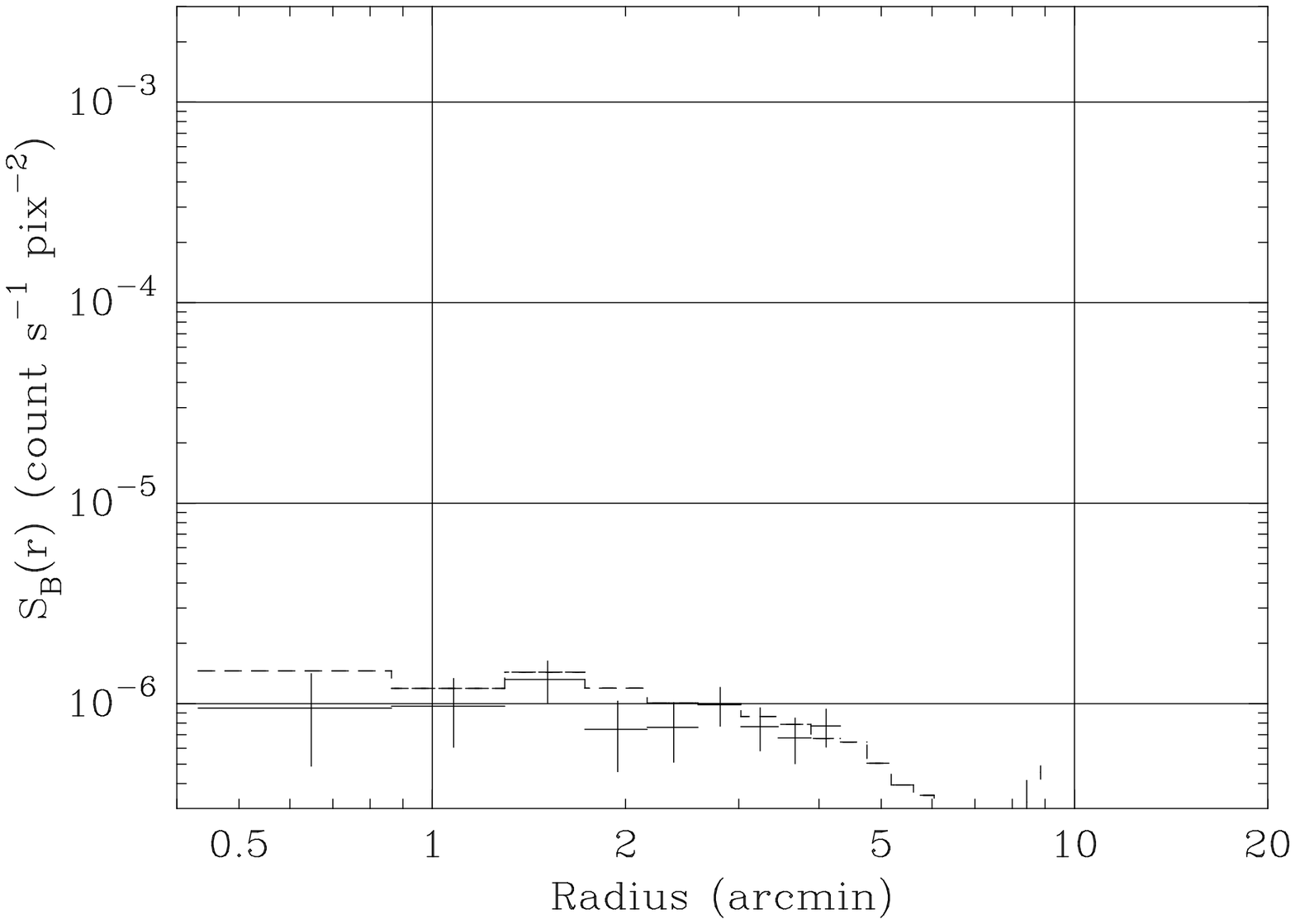}}
\end{minipage}
\caption[]{Radial count rate profiles of the HCG 57 for the GIS
 (left) and SIS (right), together with the best-fit $\beta$ model
 ($\beta=0.4$ and $R_c=3^{\prime}.75$).}
\end{figure}

\begin{figure}[hb]
\centerline{\FigureFile(95mm,95mm){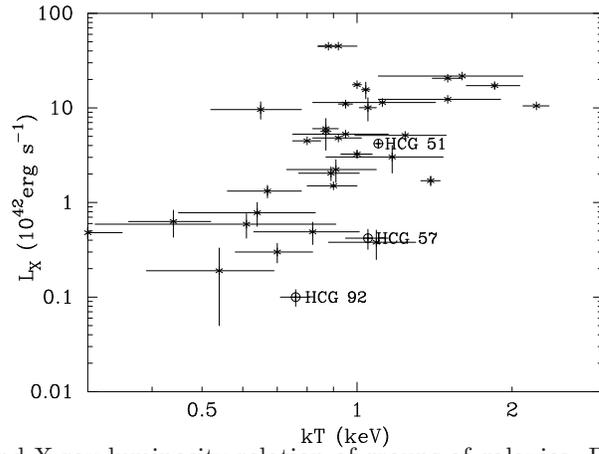}}
\caption[]{Temperature and X-ray luminosity relation of groups of
 galaxies. Data besides HCG 57, HCG 51, and HCG 92 are taken 
from Xue and Wu (2000).}
\end{figure}

\end{document}